\documentclass{article}
\usepackage{spconf,amsmath,graphicx,multirow}
\usepackage{comment}
\usepackage{amsfonts}
\usepackage[table]{xcolor}
\usepackage{enumerate}
\usepackage{diagbox}
\usepackage{enumitem}
\usepackage[super]{nth}
\usepackage[ruled,vlined]{algorithm2e}
\usepackage{wrapfig}
\usepackage{caption}
\usepackage{subcaption}
\usepackage{hyperref}       
\usepackage{url}            
\usepackage{booktabs}       

\title{Speech Representation Learning through 
\\ Self-supervised Pretraining and Multi-task Finetuning}
\name{Yi-Chen Chen$^1$,
Shu-wen Yang$^1$, Cheng-Kuang Lee$^2$, Simon See$^2$, Hung-yi Lee$^1$}
\address{$^1$National Taiwan University, Taiwan\ \ \ $^2$NVIDIA AI Technology Center, NVIDIA\\
\texttt{$^1$\{f06942069,r08944041,hungyilee\}@ntu.edu.tw}\ \ \ \texttt{$^2$\{ckl,ssee\}@nvidia.com}}

\begin{document}
\ninept
\maketitle
\begin{abstract}
Speech representation learning plays a vital role in speech processing. Among them, self-supervised learning (SSL) has become an important research direction. It has been shown that an SSL pretraining model can achieve excellent performance in various downstream tasks of speech processing.
On the other hand, supervised multi-task learning (MTL) is another representation learning paradigm, which has been proven effective in computer vision (CV) and natural language processing (NLP). However, there is no systematic research on the general representation learning model trained by supervised MTL in speech processing.
In this paper, we show that MTL finetuning can further improve SSL pretraining.
We analyze the generalizability of supervised MTL finetuning to examine if the speech representation learned by MTL finetuning can generalize to unseen new tasks.
\end{abstract}
\section{Introduction}
\label{sec:intro}

Recently many SSL approaches have been proposed for pretraining models for speech processing tasks, including pretraining with generative losses \cite{apc1,mockingjay,liu2021tera,decoar2} or discriminative losses \cite{oord2018representation,wav2vec,vq_wav2vec,wav2vec2,hsu2021hubert}. 
Some approaches use multi-task learning with multiple SSL objectives \cite{pase,pase+}.
After a shared model is pretrained with SSL to extract general representations, it can then be specialized on downstream tasks with task-specific head models and simple finetuning.
This method achieves state-of-the-art performance in many applications.

To fairly evaluate the generalizability of SSL approaches without further heavy downstream task-specific finetuning, the SUPERB benchmark \cite{yang21c_interspeech} is proposed. 
The SUPERB benchmark measures the performance of a shared model across a wide range of speech processing tasks without heavy finetuning. 
Ten tasks are included to investigate four aspects of speech: content, speaker, semantics, and paralinguistics. 
To evaluate a general model trained with SSL, the pretrained model parameters are frozen, and the fixed representations are extracted and fed into each task-specific prediction head (small downstream model) for training.
During the evaluation, the pretrained shared model and trained prediction heads are used on all tasks.
In the above scenario, some self-supervised models show outstanding performances on all the ten tasks in SUPERB. 

Supervised multi-task learning (MTL) is to train a shared model on various downstream tasks~\cite{kendall2018multi,chen2018gradnorm,sener2018multi,yu2020gradient,wang2020gradient}.  
This paper wants to investigate if MTL on various downstream tasks can further improve the representations from SSL.
In CV \cite{silberman2012indoor,everingham2010pascal} and NLP \cite{mccann2018natural,wang2018glue}, general models trained by MTL approaches can be evaluated on benchmarks that include various tasks.
However, in speech, there has not been a systematic study of general representation learning models trained by MTL of various speech processing tasks.
SpeechNet \cite{chen2021speechnet} proposes a general modularized model to perform a variety of speech processing tasks, but the purpose of SpeechNet is to allow different tasks to utilize different subsets of the general modules in SpeechNet, rather than proposing one shared model to extract general representations for training task-specific heads of downstream tasks.

In this paper, we investigate two MTL training scenarios and also one task transfer learning scenario. 
For the two MTL scenarios, we select a state-of-the-art SSL pretrained shared model in the SUPERB benchmark as the starting point for MTL. 
Then we train the shared model with MTL in two different scenarios:
\begin{itemize}
    \item All-task MTL Finetuning: Finetune the SSL pretrained shared model with all tasks in SUPERB. 
    It serves as a strong baseline for SSL approaches and the following scenarios.
    \item Leave-one-out MTL Finetuning: Finetune the SSL pretrained shared model with all but one tasks in SUPERB. 
    We can observe the influence of removing one task on the learned representations and their performance on the other tasks. 
\end{itemize}
To further examine if the representations learned with supervised MTL can generalize to an unseen new task, we have an additional Task Transfer Learning scenario.
\begin{itemize}
    \item We take a shared model from the Leave-one-out MTL scenario and freeze its parameters. We extract the representations with this shared model for training the prediction head of the remaining task that is not involved in MTL finetuning.
\end{itemize}
Through the different training scenarios above, we perform a preliminary study of the generalizability of representation learning by MTL of various speech processing tasks on a standard benchmark. 
The code is released for reproduction and future extension\footnote{https://github.com/s3prl/s3prl/tree/multi-task-distributed}.

\begin{figure*}[t]
  \centering
  \centerline{\includegraphics[width=\linewidth]{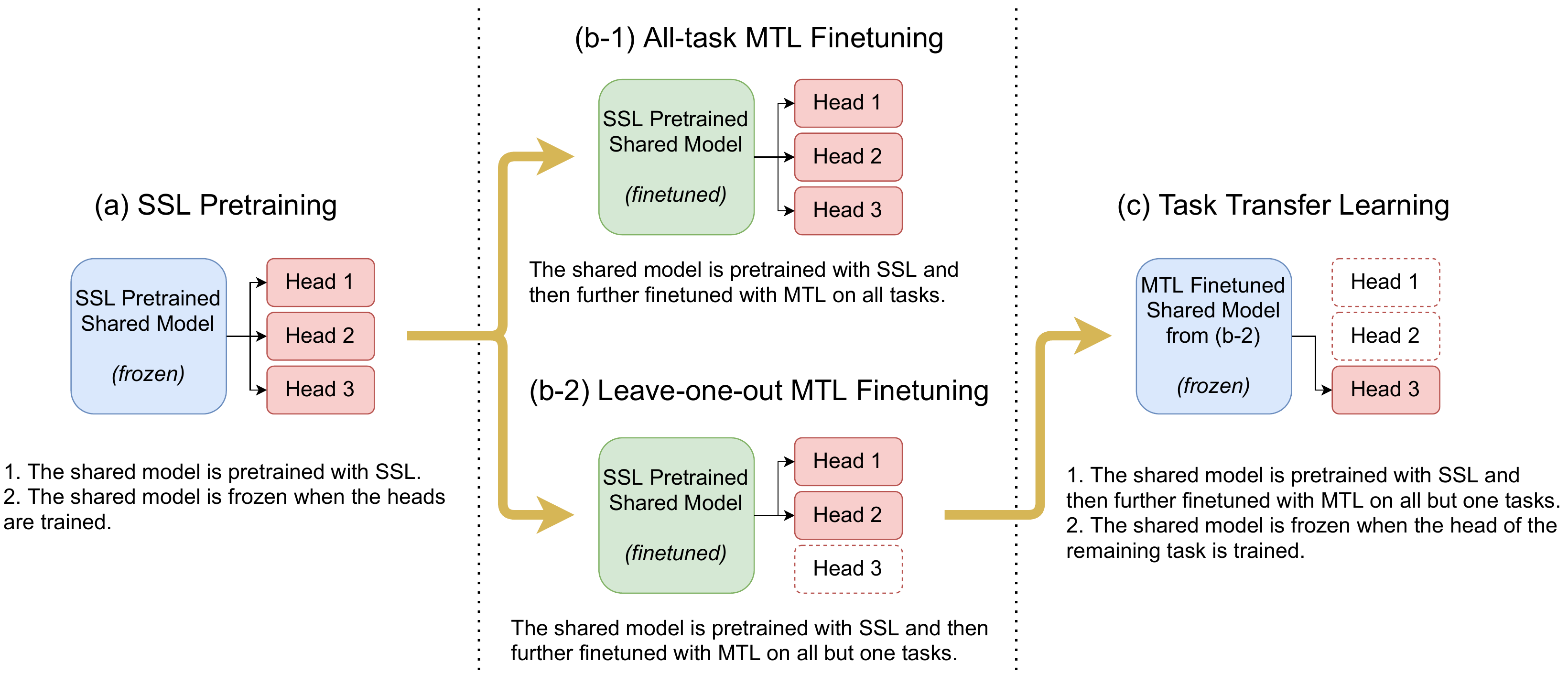}}
\caption{Four different training scenarios. Scenarios (b-1) and (b-2) require the pretrained shared model from (a). Scenario (c) requires the finetuned shared model from (b-2). The parameters of the shared model in scenarios (a) and (c) are frozen.}
\label{fig:4_scenarios}
\vspace{-0.3cm}
\end{figure*}

\section{Training Scenarios}
\label{sec:training_scenarios}
In this section, we describe four related training scenarios in the following subsections: SSL Pretraining, All-task MTL Finetuning, Leave-one-out MTL Finetuning, and Task Transfer Learning.
The two MTL finetuning scenarios require the SSL pretrained shared model, and the Task Transfer Learning scenario requires the shared model from the Leave-one-out MTL Finetuning scenario.

\subsection{SSL Pretraining}
\label{subsec:ssl_pretraining}
Many SSL approaches are evaluated and compared on the ten tasks in SUPERB \cite{yang21c_interspeech}. 
For each SSL approach, we first pretrain a model with SSL objectives.
Then we use this pretrained model as the shared model to extract representations for all downstream tasks. 
The parameters of the pretrained model are frozen. 
Then we train each task-specific prediction head (small downstream model) with the fixed representations, as illustrated in Figure \ref{fig:4_scenarios} (a).

\subsection{All-task MTL Finetuning}
\label{subsec:mtl_all_tasks}
We take the shared model pretrained in Subsection \ref{subsec:ssl_pretraining} as the starting point. 
Then we further finetune the shared model with MTL by jointly training it with downstream task-specific heads of all tasks in SUPERB, as illustrated in Figure \ref{fig:4_scenarios} (b-1). 
In this way, the shared model can be updated by the gradients of all tasks to fit the respective objectives of each task. 
Therefore, the representations extracted from the shared model can perform well on the tasks involved in MTL.
It serves as a strong baseline for SSL approaches and the following scenarios.
To further examine the generalizability of representation learning by supervised MTL, we have two additional training scenarios below.

\subsection{Leave-one-out MTL Finetuning}
\label{subsec:mtl_all_but_one_tasks}
Similarly, we take the shared model pretrained in Subsection \ref{subsec:ssl_pretraining} as the starting point.
Then we further finetune the shared model with MTL by jointly training it with downstream task-specific heads of \textbf{all but one} tasks in SUPERB, as illustrated in Figure \ref{fig:4_scenarios} (b-2).
Compared to the finetuned shared model in Subsection \ref{subsec:mtl_all_tasks}, we can observe the influence of removing one task on the learned representations and their performance on the other tasks.

\subsection{Task Transfer Learning}
\label{subsec:frozen_mtl_one_task}

We take the finetuned shared model in Subsection~\ref{subsec:mtl_all_but_one_tasks} as the pretrained shared model in this scenario. 
Then we freeze the shared model, and only train the downstream head model of the remaining task that is not used in MTL finetuning, as illustrated in Figure \ref{fig:4_scenarios} (c). 

Comparing the representations in this scenario with those learned with only SSL approaches in Subsection \ref{subsec:ssl_pretraining}, we can observe the generalizability of the representations learned with MTL on a new task compared to SSL only.
On the other hand, in comparison with the representations learned with All-task MTL in Subsection \ref{subsec:mtl_all_tasks}, we can observe how the performance of a task is influenced if this task is not involved in the MTL finetuning.

\section{Experimental Setup}
\label{sec:exp_setup}

\begin{table*}
\caption{Experimental Results of training scenarios described in Section \ref{sec:training_scenarios}. We have the numbers in Scenario (b-2) be bold or underlined if they are better or worse than (b-1) respectively.}
\label{table:exp_results}
\centering
\begin{tabular}{c|c||c|c|c|c|c|c|c|c|c|c}
\hline

\multicolumn{1}{c|}{\multirow{2}{*}{Scenario}} & Tasks for MTL & \textbf{ASR} & \textbf{PR} & \multicolumn{2}{c|}{\textbf{SF}} & \textbf{SD} & \textbf{ER} & \textbf{IC} & \textbf{KS} & \textbf{ASV} & \textbf{SID} \\ \cline{3-12}

 & Finetuning & \textbf{WER$\downarrow$} & \textbf{PER$\downarrow$} & \textbf{F1$\uparrow$} & \textbf{CER$\downarrow$} & \textbf{DER$\downarrow$} & \textbf{ACC$\uparrow$} & \textbf{ACC$\uparrow$} & \textbf{ACC$\uparrow$} & \textbf{EER$\downarrow$} & \textbf{ACC$\uparrow$} \\ \hline 
 \hline

\multirow{1}{*}{(a) SSL} & N/A & 6.42 & 5.41 & 88.53 & 25.20 & 5.88 & 64.24 & 98.34 & 96.30 & 5.11 & 81.42 \\ 
\hline
 
\multirow{1}{*}{(b-1) SSL+MTL} & all & 6.22 & 3.61 & 87.56 & 26.76 & 4.93 & 67.28 & 99.60 & 97.34 & 6.76 & 90.86 \\ 
\hline

\multirow{9}{*}{(b-2): SSL+MTL} & all but ASR & X & \underline{3.63} & \underline{87.28} & \underline{27.11} & \textbf{4.89} & \underline{65.07} & \textbf{99.63} & \textbf{97.57} & \underline{7.78} & \underline{90.69} \\
 & all but PR & \underline{6.79} & X & \underline{86.94} & \underline{27.66} & \textbf{4.81} & \underline{66.73} & \textbf{99.66} & \textbf{97.44} & \underline{7.94} & \textbf{91.16} \\
 & all but SF & \textbf{6.10} & \textbf{3.39} & X & X & \textbf{4.73} & \underline{65.71} & \underline{99.58} & \underline{97.18} & \underline{7.61} & \underline{90.70} \\
 & all but SD & \underline{6.28} & \textbf{3.54} & \textbf{87.94} & \textbf{26.31} & X & \underline{66.73} & \textbf{99.63} & \underline{97.11} & \underline{7.49} & \underline{90.79} \\
 & all but ER & \textbf{6.17} & \textbf{3.40} & \underline{87.45} & \underline{26.90} & \textbf{4.77} & X & \underline{99.55} & \underline{97.27} & \underline{7.19} & \underline{90.51} \\
 & all but IC & \textbf{6.13} & \textbf{3.34} & \textbf{87.65} & \underline{26.94} & \textbf{4.78} & \underline{66.08} & X & \underline{97.27} & \textbf{6.74} & \underline{90.55} \\
 & all but KS & \textbf{6.17} & \textbf{3.55} & \textbf{87.83} & \underline{26.88} & \textbf{4.91} & \underline{66.27} & \textbf{99.71} & X & \underline{7.86} & \underline{90.67} \\
 & all but ASV & \textbf{5.90} & \textbf{2.79} & \textbf{87.88} & \textbf{26.52} & \textbf{3.61} & \underline{64.88} & \underline{99.58} & \textbf{97.44} & X & \underline{85.06} \\
 & all but SID & \textbf{5.95} & \textbf{3.25} & \underline{87.33} & \underline{27.39} & \textbf{4.50} & \textbf{68.66} & \underline{99.55} & \underline{97.27} & \underline{9.00} & X \\ 
 \hline
 
\multirow{1}{*}{(c) Task Transfer} & N/A & 6.27 & 5.79 & 88.14 & 26.24 & 5.80 & 64.24 & 97.42 & 96.33 & 7.55 & 62.05 \\ 
\hline

\end{tabular}
\end{table*}

\begin{figure*}[t]
\begin{minipage}[t]{0.45\linewidth}
  \centering
  \centerline{\includegraphics[width=\linewidth]{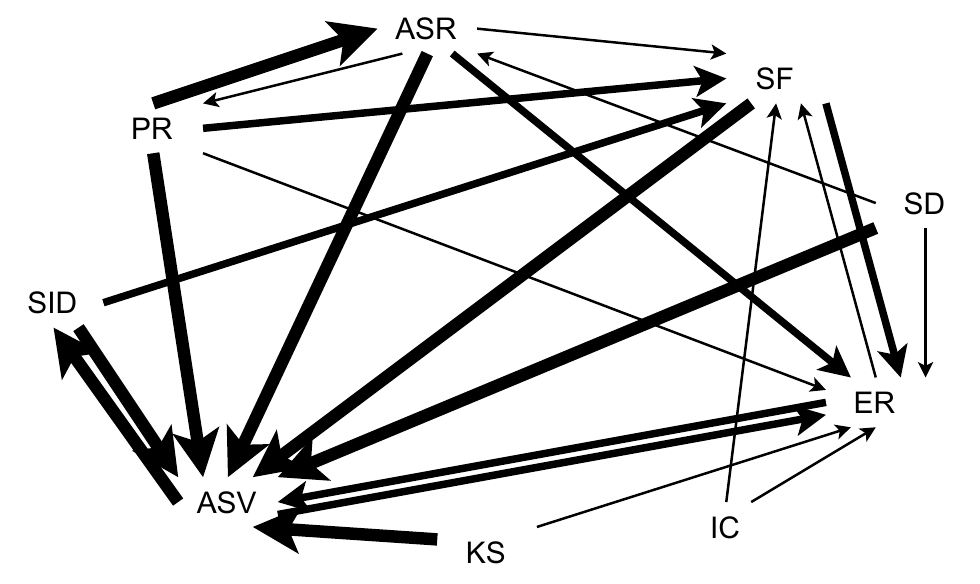}}
  \vspace{-0.2cm}
  \caption*{The improvement relations of tasks.}\medskip
  \label{minipage:improvement} 
\end{minipage}
\hspace*{\fill}\vspace{-0.1cm}
\begin{minipage}[t]{0.42\linewidth}
  \centering
  \centerline{\includegraphics[width=\linewidth]{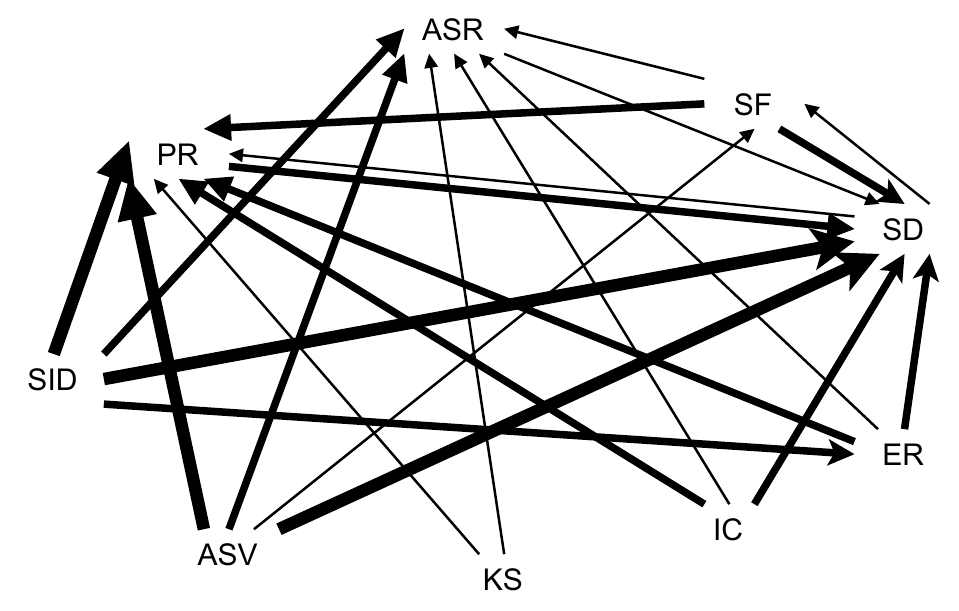}}
  \vspace{-0.2cm}
  \caption*{The hurt relations of tasks.}\medskip
  \label{minipage:hurt} 
\end{minipage}
\vspace{-0.2cm}
\caption{Two relation graphs of tasks. If a task A performs worse/better after removing a task B in MTL, we connect an edge from B to A in the improvement/hurt graphs, indicating B can improves/hurts the performance of A. The width of an edge is thin/medium/thick if the relative score change is in the range [0.5\%, 2\%), [2\%, 8\%), or larger than 8\%. If the the relative score change is less than 0.5\%, we consider it negligible and no edge is connected.}
\label{fig:relation_graphs}
\vspace{-0.3cm}
\end{figure*}

\subsection{Tasks In SUPERB}
\label{subsec:ten_tasks}

Ten tasks in SUPERB can be used to investigate four aspects of speech: \textbf{content} (Phoneme Recognition (PR), Automatic Speech Recognition (ASR), Keyword Spotting (KS), and Query by Example Spoken Term Detection (QbE)), \textbf{speaker} (Speaker Identification (SID), Automatic Speaker Verification (ASV), and Speaker Diarization (SD)), \textbf{semantics} (Intent Classification (IC) and Slot Filling (SF)), and \textbf{paralinguistics} (Emotion Recognition (ER)).
Since no downstream model training is required in QbE, we only perform MTL experiments and compare the results on the other nine tasks.

\begin{itemize}
    \item \textbf{PR} converts an utterance into a sequence of phonemes. 
    Alignment modeling is included in the PR task to avoid the potential inaccurate forced alignment.
    The evaluation metric is the phone error rate (PER).
    \item \textbf{ASR} transcribes an utterance into a sequence of words.
    While PR analyzes the performance of modeling phonetics, ASR reflects the performance of recognizing more common text units in a real-world scenario.
    The evaluation metric is the word error rate (WER).
    \item \textbf{KS} identifies preregistered keywords in an utterance by classifying the utterance into a predefined set of words.
    The task is important for on-device speech processing and requires low response time.
    The evaluation metric is the accuracy (ACC).
    \item \textbf{SID} classifies the speaker identity of an utterance in a multi-class classification setting, where the set of speakers are the same for both training and testing.
    The evaluation metric is the accuracy (ACC).
    \item \textbf{ASV} verifies whether the speakers of a pair of utterances match in a binary classification setting. Different from SID, the speakers in the testing set may not appear in the training set.
    Therefore, ASV is more challenging than SID.
    The evaluation metric is the equal error rate (EER).
    \item \textbf{SD} segments an utterance and classifies the segments into speaker identities, i.e., \textit{who is speaking when}. 
    Multiple speakers can speak simultaneously.
    Rich and various speaker characteristics should be encoded in the extracted representations for each frame to represent mixtures of signals.
    The evaluation metric is the diarization error rate (DER).
    \item \textbf{IC} classifies an utterance into predefined classes of speaker intents.
    The evaluation metric is the accuracy (ACC).
    \item \textbf{SF} converts an utterance into a sequence of semantic slot-type classes. For example, \textit{FromLocation} can be a slot-type for a spoken word \textit{Taipei}, which is known as a slot-value.
    Both slot-types and slot-values are essential for an SLU system.
    Therefore, we use two evaluation metrics for slot-types and slot-values respectively: the slot-type F1 score (F1) and the slot-value character error rate (CER).
    \item \textbf{ER} predicts an emotion class for each utterance.
    The evaluation metric is accuracy (ACC).
\end{itemize}

As for the datasets and splits used for each task, we follow the original settings in SUPERB (PR \cite{librispeech}, ASR \cite{librispeech}, KS \cite{speech_commands}, SID \cite{voxceleb1}, ASV \cite{voxceleb1}, SD \cite{cosentino2020librimix}, IC \cite{lugosch2019speech}, SF \cite{lai2020semi}, and ER \cite{iemocap}).

\subsection{The SSL Pretraining Approach In Experiments}
\label{subsec:ssl}

Many SSL approaches are evaluated and compared on the tasks in SUPERB.
Among them, HuBERT \cite{hsu2021hubert} achieves the overall best performance.
Therefore, we select HuBERT as the SSL pretraining approach across all of our experiments.

HuBERT utilizes an offline clustering algorithm on hidden representations to provide aligned target labels for a BERT-like \cite{bert} prediction.
The clustered labels at the masked locations serve as the prediction targets.
We use a weighted sum of hidden representations of all layers in the HuBERT model as the representations for downstream heads, as in SUPERB.

\subsection{Model Architecture and Implementation Details}
\label{model_arch_and_implementation}

Since MTL requires more computational resources than single-task training, we adopt HuBERT Base rather than Large in SUPERB as our shared model architecture.
For task-specific head architectures, we simply follow the settings in SUPERB.
We use a batch size of 2 for MTL finetuning in training scenarios described in Subsections \ref{subsec:mtl_all_tasks} and \ref{subsec:mtl_all_but_one_tasks}, and a batch size of 8 for downstream head training in the training scenario described in Subsection \ref{subsec:frozen_mtl_one_task}.
Each model is trained with an Adam optimizer with a linearly warmup learning rate from 0 to 1e-5 for the first 5000 steps and then a linearly decaying learning rate to 0 for 195000 steps.
For more implementation details, please refer to the released code.

\section{Experimental Results}
\label{sec:exp_results}

All experimental results are presented in Table \ref{table:exp_results}.
An up-/down-arrow beside an evaluation metric means that better performance results in a higher/lower number of that metric.
The results are grouped according to four different training scenarios corresponding to the subsections in Section \ref{sec:training_scenarios} respectively.

\subsection{The Performance Of All-task MTL Finetuning}
\label{subsec:exp_a_b_comparison}

From the comparison between Scenario (a) and Scenario (b-1), we observe that the performance of the shared model finetuned with All-task MTL is better in the tasks ASR, PR, SD, ER, IC, KS, SID, and worse only in the tasks SF and ASV than SSL pretraining.
It indicates that MTL is a strong baseline for SSL pretraining or other representation learning approaches.

One thing worth noting is that all tasks except for ASV do not suffer from overfitting in terms of the validation scores during training. 
However, the model is prone to overfit on ASV.
In Scenario (a), the model checkpoint of a downstream head can be determined by the best validation score during training for each task respectively.
Yet in Scenario (b-1), since the shared model is jointly trained with the downstream heads of all tasks, it is hard to select the model checkpoint based on the validation scores of all tasks.
In this paper, we simply select the last model checkpoint after 200,000 training steps for all tasks.
It may be a reason for the worse performance of MTL finetuning on ASV.
We leave exploring a better method to select the model checkpoint with MTL in future work.

\subsection{The Influence Of Removing One Task In MTL Finetuning}
\label{subsec:exp_b_c_comparison}

The rows in Scenario (b-2) show the results of finetuning the shared model with Leave-one-out MTL.
To compare these results with Scenario (b-1) more clearly, we have the numbers in Scenario (b-2) be bold or underlined if they are better or worse than (b-1) respectively.
Furthermore, we calculate the relative score increases/decreases of Scenario (b-2) compared to Scenario (b-1), and plot two relation graphs in Figure \ref{fig:relation_graphs} accordingly.
For SF, we use CER to plot the edges because the relative changes are small in terms of F1.
If a task A improves/regresses after removing another task B in MTL, it means that B can hurt/help A in MTL.

\begin{itemize}
    \item For \textbf{ASR}, PR is an important auxiliary task for ASR, and SD helps a little. ASR performs better after removing any of the other tasks.
    \item For \textbf{PR}, ASR helps in MTL while the other tasks hurt the performance of PR.
    We can observe that content recognition tasks such as ASR and PR are hurt the most by speaker recognition tasks such as ASV and SID.
    \item For \textbf{SF}, ASR, PR, ER, and SID help in MTL.
    IC and KS help a little in terms of F1 but hurt in terms of CER.
    SD and ASV hurt SF.
    \item For \textbf{SD}, all of the other tasks hurt SD.
    SD is also hurt the most by speaker recognition tasks such as ASV and SID.
    Although SD, ASV and SID are all related to speaker characteristics, SID and ASV focus on the utterance-level embedding, while SD aims to distinguish frame-level speaker characteristics.
    Therefore, the fine-grained information needed by SD may be lost when jointly trained with ASV or SID.
    \item For \textbf{ER}, all of the other tasks except for SID help ER.
    \item For \textbf{IC} and \textbf{KS}, the influences of the other tasks are negligible.
    \item For \textbf{ASV}, all of the other tasks except for IC help ASV.
    As discussed in Subsection \ref{subsec:exp_a_b_comparison}, ASV suffers from severe overfitting.
    Therefore, jointly learning ASV with other tasks can mitigate this issue, especially with SID.
    \item For \textbf{SID}, all of the other tasks except for PR help SID.
    ASV especially helps a lot while the others help a little.
\end{itemize}

From another perspective, the results may provide a different insight in addition to the evaluation of supervised MTL as a representation learning.
If we focus on a certain primary task, we may select proper auxiliary tasks to assist the primary task based on these MTL experimental results.
For example, if we want to have a better performance on ER, we can jointly train ER with all of the other tasks except for SID.

The transitive relations of tasks in MTL need to be further verified.
For example, suppose we know jointly training a task A with another task B can improve the performance of A; and we also know jointly training A with another task C can improve the performance of A.
Yet, we cannot conclude that jointly training A with B and C simultaneously can improve A's performance.
We leave more in-depth research of MTL and its optimization on speech processing tasks in future work.

\subsection{The Performance Of Task Transfer Learning}
\label{subec:exp_d_comparison}

To further examine the generalizability of representation learning by MTL, we present the results of the Task Transfer Learning scenario in Table \ref{table:exp_results} (c).
Compared to Table \ref{table:exp_results} (a), all of the tasks except for ASR and KS in Scenario (c) perform worse.
It indicates SSL pretraining is still a more generalizable representation learning approach for a new downstream task.
On the other hand, compared to Table \ref{table:exp_results} (b-1), all of the tasks except for SF in Scenario (c) perform worse.
It indicates whether a task is involved in MTL is crucial to the performance of this task.

To obtain better generalizability, it is worth trying to train the shared model with both SSL and MTL simultaneously as semi-supervised MTL representation learning.
We leave this exploration in future work.

\section{Conclusion And Discussion}
\label{sec:conclusion_and_discussion}

In this work, we investigate different training scenarios of supervised MTL as a speech representation learning approach along with SSL pretraining on a benchmark with various speech processing tasks.
We analyze the generalizability of representations learned with supervised MTL empirically.

This paper is only a preliminary study of MTL with various speech processing tasks.
The performance of MTL is dependent on many factors, such as the amount of data, task relationships, noise and so on. 
These factors should be isolated and investigated with more theoretical analyses and empirical experiments in the future.


\bibliographystyle{IEEEbib}

\begin{thebibliography}{10}

\bibitem{apc1}
Yu-An Chung, Wei-Ning Hsu, Hao Tang, and James Glass,
\newblock ``{An Unsupervised Autoregressive Model for Speech Representation
  Learning},''
\newblock in {\em Interspeech}, 2019, pp. 146--150.

\bibitem{mockingjay}
Andy~T. Liu, Shu-wen Yang, Po-Han Chi, Po-chun Hsu, and Hung-yi Lee,
\newblock ``Mockingjay: Unsupervised speech representation learning with deep
  bidirectional transformer encoders,''
\newblock {\em ICASSP}, 2020.

\bibitem{liu2021tera}
Andy~T Liu, Shang-Wen Li, and Hung-yi Lee,
\newblock ``Tera: Self-supervised learning of transformer encoder
  representation for speech,''
\newblock {\em IEEE/ACM Transactions on Audio, Speech, and Language
  Processing}, vol. 29, pp. 2351--2366, 2021.

\bibitem{decoar2}
Shaoshi Ling and Yuzong Liu,
\newblock ``{DeCoAR} 2.0: Deep contextualized acoustic representations with
  vector quantization,''
\newblock {\em arXiv preprint arXiv:2012.06659}, 2020.

\bibitem{oord2018representation}
Aaron van~den Oord, Yazhe Li, and Oriol Vinyals,
\newblock ``Representation learning with contrastive predictive coding,''
\newblock {\em arXiv preprint arXiv:1807.03748}, 2018.

\bibitem{wav2vec}
Steffen Schneider, Alexei Baevski, Ronan Collobert, and Michael Auli,
\newblock ``wav2vec: Unsupervised pre-training for speech recognition.,''
\newblock in {\em Interspeech}, 2019.

\bibitem{vq_wav2vec}
Alexei Baevski, Steffen Schneider, and Michael Auli,
\newblock ``vq-wav2vec: Self-supervised learning of discrete speech
  representations,''
\newblock in {\em ICLR}, 2020.

\bibitem{wav2vec2}
Alexei Baevski, Yuhao Zhou, Abdelrahman Mohamed, and Michael Auli,
\newblock ``wav2vec 2.0: {A} framework for self-supervised learning of speech
  representations,''
\newblock in {\em NeurIPS}, 2020.

\bibitem{hsu2021hubert}
Wei-Ning Hsu, Yao-Hung~Hubert Tsai, Benjamin Bolte, Ruslan Salakhutdinov, and
  Abdelrahman Mohamed,
\newblock ``Hubert: How much can a bad teacher benefit asr pre-training?,''
\newblock in {\em ICASSP}. IEEE, 2021, pp. 6533--6537.

\bibitem{pase}
Santiago Pascual, Mirco Ravanelli, Joan Serr{\`a}, Antonio Bonafonte, and
  Yoshua Bengio,
\newblock ``Learning problem-agnostic speech representations from multiple
  self-supervised tasks,''
\newblock in {\em Interspeech}, 2019, pp. 161--165.

\bibitem{pase+}
Mirco Ravanelli, Jianyuan Zhong, Santiago Pascual, Pawel Swietojanski, Joao
  Monteiro, Jan Trmal, and Yoshua Bengio,
\newblock ``Multi-task self-supervised learning for robust speech
  recognition,''
\newblock in {\em ICASSP}, 2020, pp. 6989--6993.

\bibitem{yang21c_interspeech}
Shu wen Yang, Po-Han Chi, Yung-Sung Chuang, Cheng-I~Jeff Lai, Kushal Lakhotia,
  Yist~Y. Lin, Andy~T. Liu, Jiatong Shi, Xuankai Chang, Guan-Ting Lin,
  Tzu-Hsien Huang, Wei-Cheng Tseng, Ko~tik Lee, Da-Rong Liu, Zili Huang, Shuyan
  Dong, Shang-Wen Li, Shinji Watanabe, Abdelrahman Mohamed, and Hung yi~Lee,
\newblock ``{SUPERB: Speech Processing Universal PERformance Benchmark},''
\newblock in {\em Proc. Interspeech 2021}, 2021, pp. 1194--1198.

\bibitem{kendall2018multi}
Alex Kendall, Yarin Gal, and Roberto Cipolla,
\newblock ``Multi-task learning using uncertainty to weigh losses for scene
  geometry and semantics,''
\newblock in {\em Proceedings of the IEEE conference on computer vision and
  pattern recognition}, 2018, pp. 7482--7491.

\bibitem{chen2018gradnorm}
Zhao Chen, Vijay Badrinarayanan, Chen-Yu Lee, and Andrew Rabinovich,
\newblock ``Gradnorm: Gradient normalization for adaptive loss balancing in
  deep multitask networks,''
\newblock in {\em International Conference on Machine Learning}. PMLR, 2018,
  pp. 794--803.

\bibitem{sener2018multi}
Ozan Sener and Vladlen Koltun,
\newblock ``Multi-task learning as multi-objective optimization,''
\newblock in {\em NeurIPS}, 2018, pp. 525--536.

\bibitem{yu2020gradient}
Tianhe Yu, Saurabh Kumar, Abhishek Gupta, Sergey Levine, Karol Hausman, and
  Chelsea Finn,
\newblock ``Gradient surgery for multi-task learning,''
\newblock {\em NeurIPS}, vol. 33, 2020.

\bibitem{wang2020gradient}
Zirui Wang, Yulia Tsvetkov, Orhan Firat, and Yuan Cao,
\newblock ``Gradient vaccine: Investigating and improving multi-task
  optimization in massively multilingual models,''
\newblock in {\em ICLR}, 2020.

\bibitem{silberman2012indoor}
Nathan Silberman, Derek Hoiem, Pushmeet Kohli, and Rob Fergus,
\newblock ``Indoor segmentation and support inference from rgbd images,''
\newblock in {\em European conference on computer vision}. Springer, 2012, pp.
  746--760.

\bibitem{everingham2010pascal}
Mark Everingham, Luc Van~Gool, Christopher~KI Williams, John Winn, and Andrew
  Zisserman,
\newblock ``The pascal visual object classes (voc) challenge,''
\newblock {\em International journal of computer vision}, vol. 88, no. 2, pp.
  303--338, 2010.

\bibitem{mccann2018natural}
Bryan McCann, Nitish~Shirish Keskar, Caiming Xiong, and Richard Socher,
\newblock ``The natural language decathlon: Multitask learning as question
  answering,''
\newblock {\em arXiv preprint arXiv:1806.08730}, 2018.

\bibitem{wang2018glue}
Alex Wang, Amanpreet Singh, Julian Michael, Felix Hill, Omer Levy, and Samuel~R
  Bowman,
\newblock ``Glue: A multi-task benchmark and analysis platform for natural
  language understanding,''
\newblock in {\em ICLR}, 2018.

\bibitem{chen2021speechnet}
Yi-Chen Chen, Po-Han Chi, Shu-wen Yang, Kai-Wei Chang, Jheng-hao Lin, Sung-Feng
  Huang, Da-Rong Liu, Chi-Liang Liu, Cheng-Kuang Lee, and Hung-yi Lee,
\newblock ``Speechnet: A universal modularized model for speech processing
  tasks,''
\newblock {\em arXiv preprint arXiv:2105.03070}, 2021.

\bibitem{librispeech}
V.~{Panayotov}, G.~{Chen}, D.~{Povey}, and S.~{Khudanpur},
\newblock ``Librispeech: An {ASR} corpus based on public domain audio books,''
\newblock in {\em ICASSP}, 2015, pp. 5206--5210.

\bibitem{speech_commands}
Pete Warden,
\newblock ``Speech commands: A public dataset for single-word speech
  recognition.,''
\newblock {\em Dataset available online}, 2017.

\bibitem{voxceleb1}
Arsha Nagrani, Joon~Son Chung, Weidi Xie, and Andrew Zisserman,
\newblock ``Voxceleb: Large-scale speaker verification in the wild,''
\newblock {\em Computer Speech \& Language}, vol. 60, pp. 101027, 2020.

\bibitem{cosentino2020librimix}
Joris Cosentino, Manuel Pariente, Samuele Cornell, Antoine Deleforge, and
  Emmanuel Vincent,
\newblock ``Librimix: An open-source dataset for generalizable speech
  separation,''
\newblock {\em arXiv preprint arXiv:2005.11262}, 2020.

\bibitem{lugosch2019speech}
Loren Lugosch, Mirco Ravanelli, Patrick Ignoto, Vikrant~Singh Tomar, and Yoshua
  Bengio,
\newblock ``Speech model pre-training for end-to-end spoken language
  understanding,''
\newblock in {\em Interspeech}, 2019, pp. 814--818.

\bibitem{lai2020semi}
Cheng-I Lai, Yung-Sung Chuang, Hung-Yi Lee, Shang-Wen Li, and James Glass,
\newblock ``Semi-supervised spoken language understanding via self-supervised
  speech and language model pretraining,''
\newblock in {\em ICASSP}, 2021.

\bibitem{iemocap}
Carlos Busso et~al.,
\newblock ``Iemocap: Interactive emotional dyadic motion capture database,''
\newblock {\em Language resources and evaluation}, vol. 42, no. 4, pp.
  335--359, 2008.

\bibitem{bert}
Jacob Devlin, Ming-Wei Chang, Kenton Lee, and Kristina Toutanova,
\newblock ``Bert: Pre-training of deep bidirectional transformers for language
  understanding,''
\newblock in {\em NAACL}, 2019, pp. 4171--4186.

\end{thebibliography}

\end{document}